          [2001/04/25 1.1 (PWD)]
\documentclass[a4paper,twocolumn]{esapub2005} 
\usepackage{times}
\usepackage{graphicx}
\usepackage{natbib}

\title{IGR~J17098-3628: an X-ray Nova discovered by INTEGRAL}
\author[1]{S. A. Grebenev}
\author[1]{S. V. Molkov}
\affil[1]{Space Research Institute, Russian Academy of Sciences,
Profsoyuznaya 84/32, Moscow 117997, Russia}
\author[1,2]{M. G. Revnivtsev}
\author[1,2]{R. A. Sunyaev}
\affil[2]{Max-Planck-Institut f\"ur Astrophysik,
Karl-Schwarzschild-Str. 1, Garching D-85741, Germany}

\def\arcmin{\hbox{$^\prime$}}
\def\arcsec{\hbox{$^{\prime\prime}$}}
\def\la{\mathrel{\hbox{\rlap{\hbox{\lower4pt\hbox{$\sim$}}}\hbox{$<$}}}}
\def\ga{\mathrel{\hbox{\rlap{\hbox{\lower4pt\hbox{$\sim$}}}\hbox{$>$}}}}
\def\uh{\hbox{$^{\rm h}$}}
\def\um{\hbox{$^{\rm m}$}}
\def\us{\hbox{$^{\rm s}$}}
\def\fsec{\hbox{$.\!\!^{\rm s}$}}
\def\fmin{\hbox{$.\!\!^{\rm m}$}}
\def\farcs{\hbox{$.\!\!^{\prime\prime}$}}
\def\farcm{\hbox{$.\!^{\prime}$}}
\def\deg{\hbox{$^\circ$}}

\begin{document}


\maketitle

\begin{abstract}
We report the discovery with INTEGRAL on March 24, 2005, and
follow-up observations of the distant Galactic X-ray nova
IGR\,J17098-3628.

\end{abstract}

\section{Introduction}

X-ray novae are bright transients flared up on the sky for
several months due to unsteady accretion of matter from a low
mass companion onto a compact object. Doppler spectroscopy of
optical lines from such systems performed after their switching
off in X-rays allowed their mass functions to be measured and
indicates that the compact objects in these systems are black
holes. The canonical X-ray novae have very soft black body
spectra near their brightness maximum (the reason to call them
sometimes as soft X-ray transients) and very hard (extended to
$\sim200$ keV) Comptonized spectra at the short initial and long
decaying phases of the outburst. It is obvious that the spectral
evolution of X-ray novae should trace changes in regimes of disc
accretion onto a black hole in these sources connected with
changes in the accretion rate. This opportunity for studying the
accretion regimes and the unique opportunity for finding new
black holes and measuring their mass make X-ray novae very
interesting targets for observers and astrophysicists.

Long term X-ray observations by GINGA, MIR/KVANT, GRANAT, CGRO
and RXTE indicate that X-ray novae flare up approximately once a
year that allowed us to believe that the {\it International
Gamma-Ray Astrophysics Laboratory\/} (INTEGRAL)
\citep{winkler03} with its wide fields of view of main
telescopes, high sensitivity in the hard X-ray band and
excellent capabilities for broad band spectroscopy will be able
to detect and investigate many X-ray novae and provide us with
new discoveries related to their peculiar X-ray
properties. However, two years of the INTEGRAL nominal operation
have passed, more than one hundred of new hard X-ray sources
have been discovered, several recurrent transients containing a
black hole (XTE\,J1550-564, 4U\,1630-472,
H\,1743-322/IGR\,J17464-3213, SLX\,1746-331, \mbox{GX\,339-4})
have flared up and been investigated but no one really new X-ray
nova has been observed. In this paper we report the discovery
with INTEGRAL probably the first source of this type,
IGR\,J17098-3628, describe its X-ray properties and some results
from the follow-up observations.
\section{Discovery and X-ray observations}

The new transient source IGR\,J17098-3628 was discovered by
\citet{grebenev05a} with the IBIS/ISGRI telescope
\citep{lebrun03} on board INTEGRAL on March 24.33-25.58, 2005
(UT), during deep Open Program observations of the Galactic
center field. The signal-to-noise (S/N) ratio for the source in
the mosaic image accumulated during this observation was
$\simeq22$ in the 18--45 keV band and $\simeq15$ in the 45--70
keV band, the corresponding average fluxes were $28.2\pm1.4$ and
$38.7\pm2.8$ mCrab. The source was variable on a time scale of
hours reaching the maximum flux levels of 60 and 95 mCrab in
these bands.

\begin{figure*}[p]
\centering
\begin{minipage}{16cm}
\hspace{3mm}\includegraphics[width=0.48\linewidth]{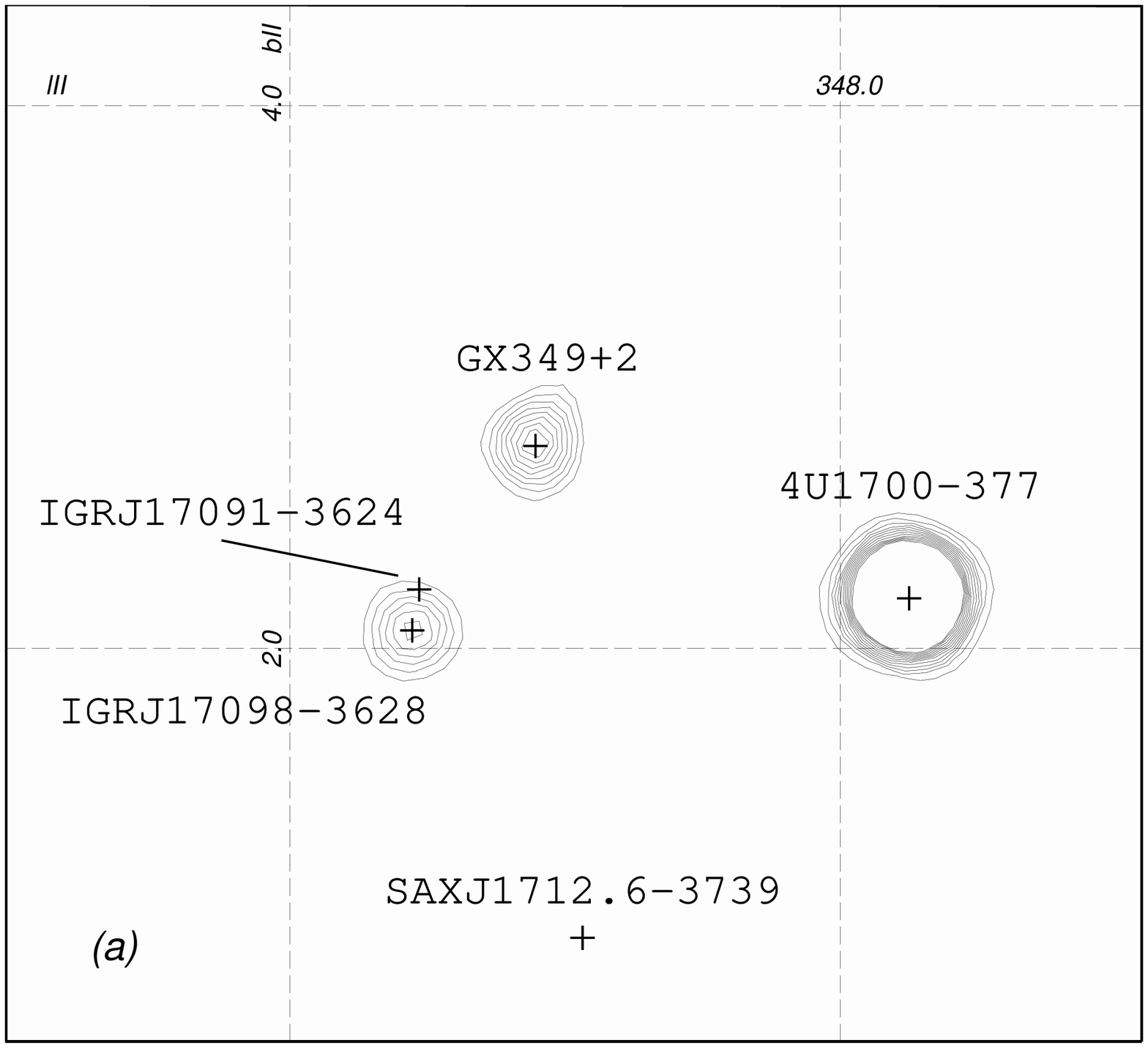}
\includegraphics[width=0.48\linewidth]{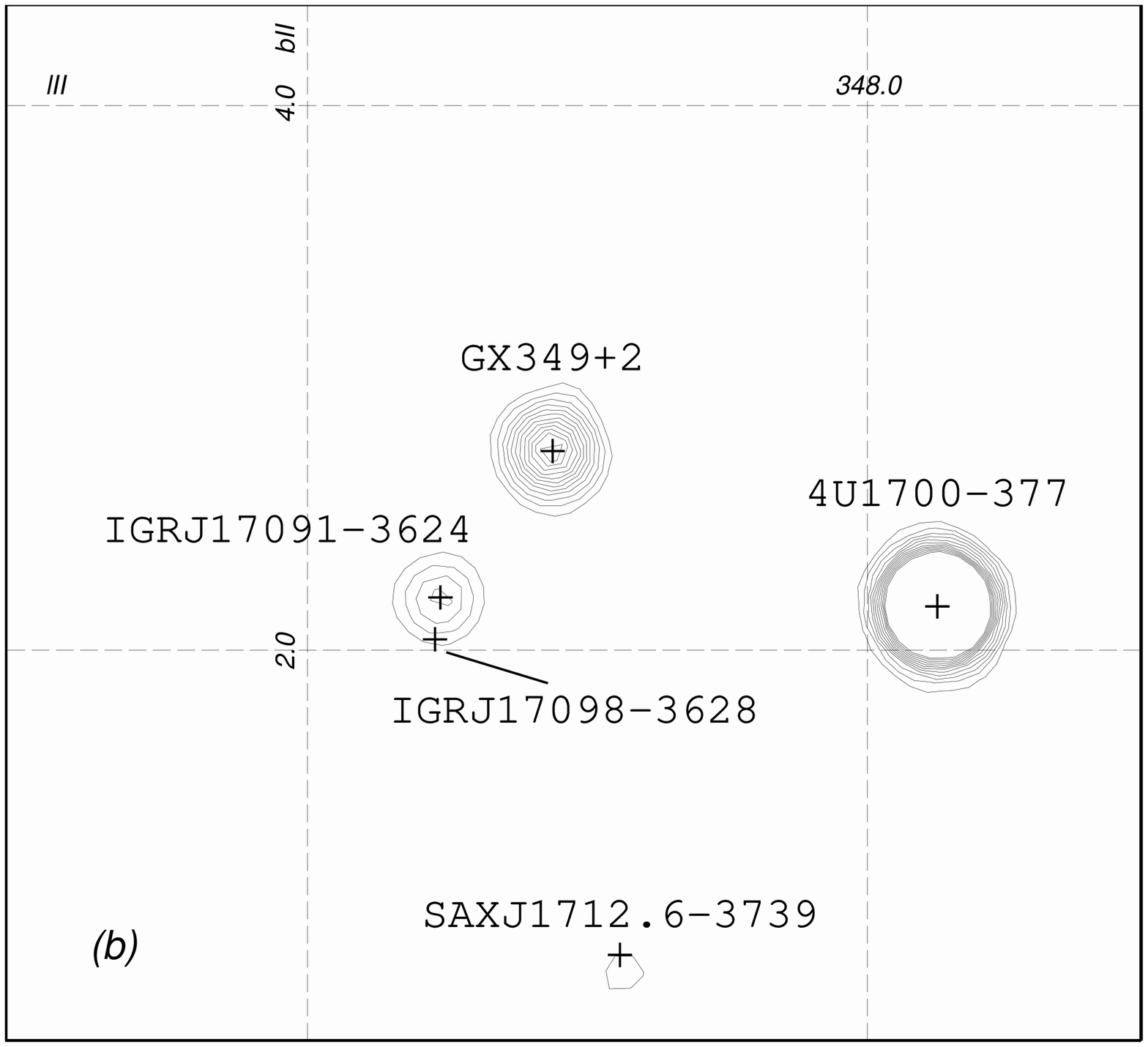}
\caption{X-ray images (S/N maps) of the region near GX\,349+2 
obtained with IBIS/ISGRI in the 18--45 keV band: (a) on March
24.33-25.58, 2005 (UT) when the outburst of the new transient
\mbox{IGR J17098-3628} was discovered, and (b) on February
16.23-18.66, 2004 (UT) when \mbox{IGR J17091-3624}, the twin of
\mbox{IGR J17098-3628}, was observed in its active state. Contours
show regions of confident detection of sources at the S/N
levels of 5, 8, 11, 14, ..., 44.\label{images}}
\vspace{1.8cm}

\includegraphics[width=0.99\linewidth]{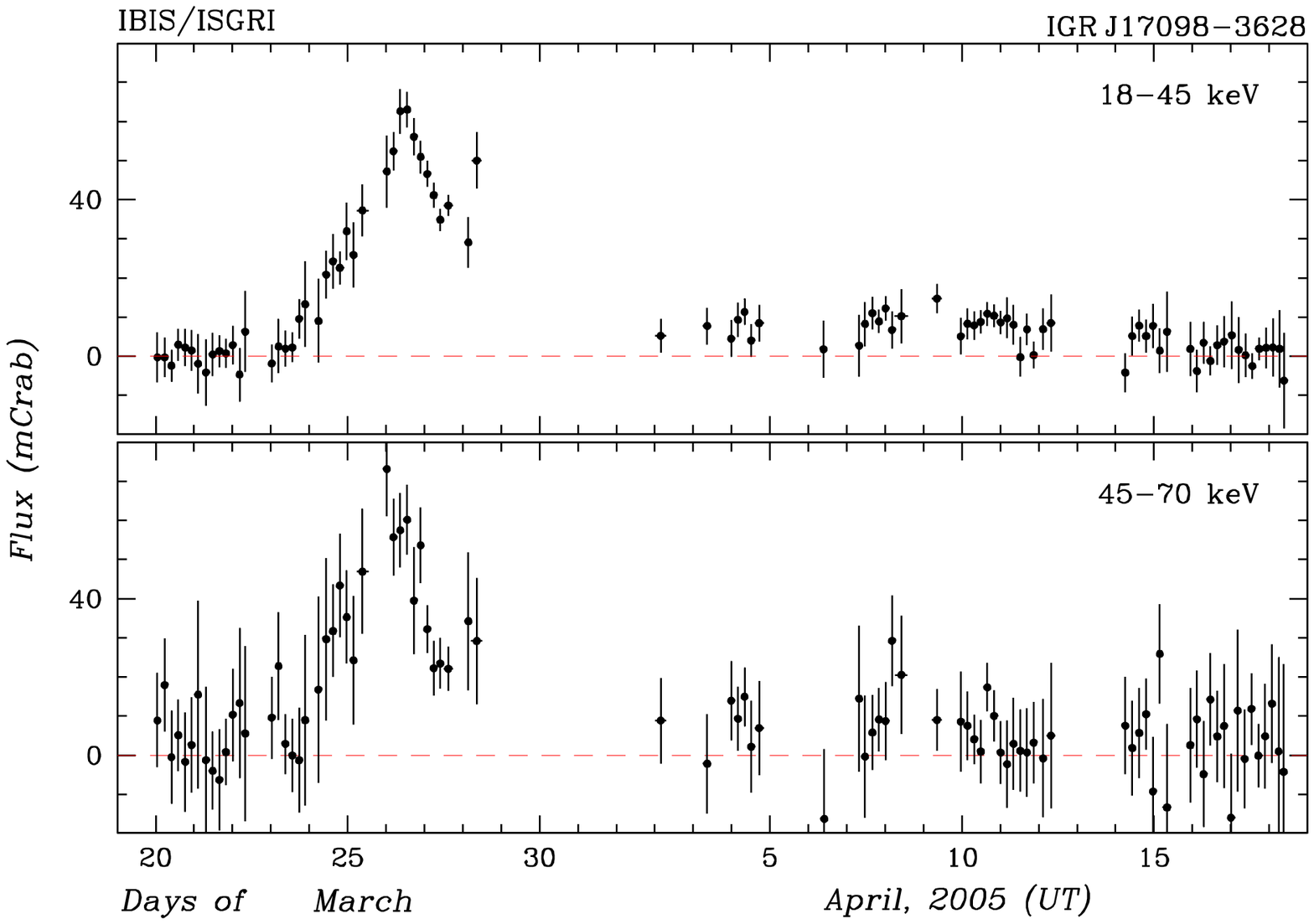}
\caption{Light curve of IGR\,J17098-3628 measured with IBIS/ISGRI
in two energy bands during the first month of its
activity. Every point is the result of averaging the fluxes
obtained in 7--8 subsequent individual pointings (covering
together an interval of $\sim$1600 s in duration). Flux of 1
mCrab corresponds to 1.1 and 0.5$\times$10$^{-11}$\/ erg cm$^{-2}$
s$^{-1}$ in the 18--45 and 45--70 keV bands for the source with
the Crab-like spectrum.\label{lcurve}}
\end{minipage}
\end{figure*}

The S/N mosaic map obtained in 18--45 keV X-rays with ISGRI
during this observation is presented in Fig.\,\ref{images}a. It
shows that IGR\,J17098-3628 was located in the close vicinity
($9.4$\arcmin\ away) of the other strongly variable INTEGRAL
source IGR\,J17091-3624 \citep{kuulkers03, mikej03, intzand03,
capitanio06}. The position of IGR\,J17098-3628 measured with
ISGRI, $R.A. = 17$\uh09\um48\us, $Decl. =
-36$\deg28\arcmin12\arcsec\ (equinox 2000.0, error radius
2\arcmin), kept it however well outside the error circle for the
position of the second source ($R.A. = 17$\uh09\um06\us,
$Decl. = -36$\deg24\arcmin07\arcsec, error radius 0.8\arcmin\
\citep{mikej03}). To illustrate the capability of the telescope
for distinguishing these sources we are giving in
Fig.\,\ref{images}b the S/N map of the same region as in
Fig.\,\ref{images}a but obtained with \mbox{ISGRI} on February
16.23-18.66, 2004, when IGR\,J17091-3624 was in the bright
state.

Two next sets of our observations of the Galactic center field
(on March 26.10-26.78 and 28.05-28.46 UT) showed that
IGR\,J17098-3628 was strongly evolving in both brightness and
spectral shape \citep{grebenev05b}. The spectra measured on March
24--25 and March 26 could be satisfactorily described by a simple
power law in the broad 18--200 keV band without any signs of a
high energy cut-off. The photon index was equal to $1.81\pm0.09$
and $2.20\pm0.06$, respectively. The spectrum measured on March
28 was significant only below 70 keV. Its approximation with a
power law led to the photon index $3.00\pm0.25$.

Figure\,\ref{lcurve} gives light curves of the source in two
energy bands, 18--45 and 45--70 keV, measured with ISGRI during
this and several subsequent weeks of observations. Data from the
INTEGRAL Galactic Bulge monitoring program (partly presented
in \citep{mowlavi05}) and several other Open Program
observations (currently public) were added here to data of our
deep view to the Galactic center field. Unfortunately there was
a gap in the data between March 29 and April 2 because of
calibration observations of the Crab nebula performed with
INTEGRAL.
The figure shows that the steady increase in the hard X-ray flux
observed during the first days of the outburst stopped on March
26 when the level of $\sim65$ mCrab has been reached. The flux
began to decrease, slowly in the 18--45 keV band and abruptly in
the 45--70 keV band, but then stabilized and was detectable at
the level of 5--10 mCrab during at least two next weeks.

The observed hard X-ray variability did not reflect changes in
the rate of energy release in \mbox{IGR\,J17098-3628}. A bulk of
the total luminosity was emitted in the softer ($\la15$ keV)
X-ray band. Measurements with the \mbox{JEM-X} monitor on board
INTEGRAL could be useful but the source was usually outside the
narrow field of view of \mbox{JEM-X} or at its very edge, in the
region of low sensitivity. The RXTE observatory carried out a
cross-scanning of this field on March 29.179-29.227 (UT) and
detected the source with the 3--20 keV flux of $\sim80$ mCrab
\citep{grebenev05b} (we thank the RXTE team for organizing such
a prompt observation). Its best-fit position
($R.A.=17$\uh09\um38\us, $Decl.=-36$\deg27\arcmin41\arcsec,
error radius 5\arcmin) was generally consistent with that of
ISGRI (being only 2\farcm1 away).
\begin{figure}[h]
\vspace{-3mm}

\hspace{-3mm}\includegraphics[width=1.05\linewidth]{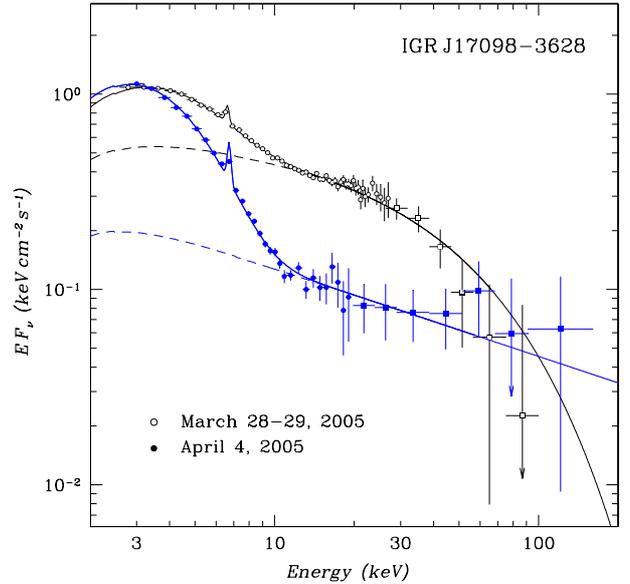}

\vspace{-2mm}
\caption{Spectral evolution of IGR\,J17098-3628 according to
INTEGRAL/ISGRI (squares) and RXTE/PCA (circles) data. Best-fit
models dominated by disk black-body spectra are shown by solid
lines, hard power law components (with a cut-off for March 28)
--- by dashed lines.\label{spec}}
\vspace{-1mm}
\end{figure}
The broad-band spectrum $E^2\,dN/dE$ of IGR\,J17098-3628
measured with RXTE/PCA (on March 29) and IBIS/ISGRI (on March
28) is shown in Fig.\,\ref{spec} by open points. The spectrum
is complex consisting of a soft black body component (we
approximated it with the {\tt diskbb} model of XSPEC) and a hard
tail (approximated with the {\tt cutoffpl} model of XSPEC).  The
inner temperature and radius of the disk were $kT_{\rm
in}=1.20\pm0.01$ keV and $R_{\rm in}(\cos{i})^{1/2}=
(6.6\pm0.1)\,d_{10}$ km, the photon index and cut-off energy of
the hard tail were $\alpha= 2.14\pm0.09$ and $E_0=46\pm9$
keV. Here $i$ is an inclination angle of the disk and $d_{10}$
is a distance to the source in units of 10 kpc. To improve the
fit an absorption with the column density $N_{\rm
H}=8\times10^{21}$ cm$^{-2}$, consistent with the Galactic
value, and a narrow ($\sigma=0.1$ keV) Gaussian line at
$6.63\pm0.08$ keV with the flux $(9.6\pm2.6)\times10^{-4}$ phot
cm$^{-2}$ s$^{-1}$ have been included. The ISGRI spectrum was
normalized to the PCA spectrum by a factor $A=1.19\pm0.14$
derived from the fit. The luminosity was $\simeq
2.4\times10^{37}\ d_{10}^2$ and $5.0\times10^{36}\
d_{10}^2$ erg s$^{-1}$ in the 3--20 and 20--200 keV bands. Note
that the hard component contributed significantly $\sim55$\% to
the 3--20 keV luminosity. The luminosity of the {\tt diskbb}
component $L_{\rm dbb}=4\pi R_{\rm in}^2\,\sigma T_{\rm
in}^4\simeq1.2\times10^{37}\ d_{10}^2\ (\cos{i})^{-1}$ erg s$^{-1}$.

Later RXTE carried our several pointed observations of the
source (with a $\sim25$\arcmin\ offset to discriminate possible
contributions from IGR\,J17091-3624). The spectrum measured on
April 4 with PCA and ISGRI is shown in Fig.\,\ref{spec} by
filled points.  The best-fit parameters were $kT_{\rm
in}=0.99\pm0.01$ keV, $R_{\rm in}(\cos{i})^{1/2}=
(12.7\pm0.2)\,d_{10}$ km, $\alpha= 2.45\pm0.12$ and $E_0>74$ keV,
$A=0.71\pm0.12$. The flux in the iron line was consistent with that
measured on March 29 confirming its interstellar (Galactic)
origin.  The luminosity was $\simeq 1.6\times10^{37}\
d_{10}^2$ (3--20 keV) and $1.5\times10^{36}\ d_{10}^2$ erg
s$^{-1}$ (20--200 keV), $L_{\rm dbb}\simeq1.9\times10^{37}
d_{10}^2(\cos{i})^{-1}$ erg s$^{-1}$.

\section{RADIO AND OPTICAL IDENTIFICATION}

To allow identification of IGR J17098-3628 in soft X-ray,
optical and radio bands we initiated its TOO observation with
the SWIFT/XRT telescope. The observation was carried out on May 1.68-1.76,
2005 (UT) with an exposure time of 2.8 ks. 
The analysis \citep{kennea05} of these data has revealed
a bright X-ray source with the coordinates $R.A. =
17$\uh09\um45\fsec9, $Decl.=-36${\deg}27{\arcmin}57\arcsec\ and
the error radius of about 5\arcsec\ (90\% containment). This
position is 30\arcsec\ from the INTEGRAL position. 
The source's average 0.5--10 keV flux 
corresponded to the luminosity $1.6\times 10^{37}\ d_{10}^2$ erg
s$^{-1}$ (non-corrected for absorption). 
The flux dropped by $\sim8$\% during the observation.

Following the SWIFT/XRT localization the possible optical/IR
counterparts for IGR\,J17098-3628, associated with 2MASS\,J17094612-3627573, have been proposed  
\citep{kong05} and 
nominally confirmed by new observations from the 6.5-m
Magellan-Baade telescope \citep{steeghs05a} and SWIFT/UVOT
\citep{blustin05}.  However, because of the low Galactic
latitude of IGR\,J17098-3628 (see Fig.\,\ref{images}), the
stellar density in this field is high and the chance of finding
a non-related star even within the narrow XRT error box was
considerable. The long term VLA observations \citep{rupen05}
have shown that 2MASS\,J17094612-3627573 did not relate to the
X-ray source.  These observations were carried out on March 31,
April 5, April 12, and May 4, all at 4.86 GHz. The first data
set showed the only significant radio source within the
2\arcmin\ INTEGRAL error circle, located at $R.A.=
17$\uh09\um45\fsec$934\pm0$\fsec011, $Decl.=
-36$\deg27\arcmin57\farcs$30\pm0$\farcs55. Its flux density on
March 31 was $0.34\pm0.07$ mJy. The later observations gave
nominal flux densities of $0.06\pm0.07$, $0.16\pm0.07$ and
$0.21\pm0.05$ mJy/beam; only the last was a detection. The
radio transient lies 0\farcs5 from the SWIFT/XRT position and
2\farcs8 from that of 2MASS J17094612-3627573. Its fading from
March 31 to April 5, and possible re-appearance around May 4,
are consistent with the X-ray evolution
\citep{grebenev05b, kennea05} and indicate that this is indeed
the radio counterpart to the X-ray transient.

\begin{figure}
\centering
\includegraphics[width=0.99\linewidth]{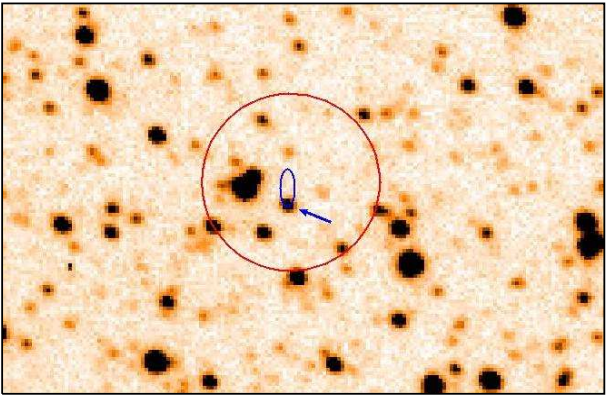}
\caption{Magellan-IMACS I-band image of the field near IGR\,J17098-3628. 
Circle and ellipse denote the 90\% error regions of the SWIFT
and VLA localization, arrow --- the variable optical
counterpart. 2MASS J17094612-3627573 with $V\simeq19.3$ is
3\arcsec\ to the left (from
\citep{steeghs05b}). \label{image_opt}}
\vspace{-4mm}
\end{figure}

\citet{steeghs05b} re-investigated their Magellan-Baade I-band
images obtained on April 9, 2005 \cite{steeghs05a} and found
(see Fig.\,\ref{image_opt}) a point source located at
$R.A.=17$\uh09\um45\fsec93, $Decl.= -36$\deg27\arcmin58\farcs2
(0\farcs2 uncertainty). This
optical position is consistent within 2-$\sigma$ error with that
derived from the radio observations. The additional I-band
images taken with the telescope on May 13.388--13.394, 2005 (UT)
indicated that the optical source has faded by
0\fmin$12\pm0$\fmin02 since April~9. The positional coincidence
with the SWIFT and VLA detections and the photometric
variability suggest that this source is indeed the optical
counterpart to IGR\,J17098-3628.

\section{Conclusions}
The observed behaviour of IGR J17098-3628 suggests that it was
an X-ray nova at an initial stage of the outburst and thus --- a
new black hole candidate. Its spectra have been successfully
described with a sum of a disk black-body emission
and a hard power law tail. Note that:\\
\makebox[5mm]{}1). The temperature $kT\sim1$ keV of the soft 
spectral component of IGR\,J17098-3628 was much smaller than
that typical of LMXBs containing neutron stars. The variable 2
keV component in such sources originates from a hot boundary
layer at the neutron star's surface.\\ 
\makebox[5mm]{}2). The value of $R_{\rm
in}$ measured on April 4 can be compared with the radius
$R_0=6GM/c^2$ of the marginal stable orbit of a black
hole. Assuming that the black hole's mass $M\geq3 M_{\odot}$ we
get the following restriction for the inclination and distance
$\cos{i}\leq0.22 d_{10}^2$ ($i\geq77$\deg\ for $d=10$ kpc,
i.e. we see the disk in this system nearly from its edge). The
{\tt diskbb} luminosity $L_{\rm dbb}\ga5.3\times10^{37}$ and
$\ga8.7\times10^{37}$erg s$^{-1}$ on March 29 and April 4.  The
small value of $R_{\rm in}$ and $L_{\rm dbb}$ measured on March
29 was probably connected with a notable contribution of the
power law component to the soft $\la15$ keV part of the spectrum
that affected the disk parameters. In reality, the hard tail
should be gradually formed from photons of the disk black body
spectrum in result of Comptonization.\\
\makebox[5mm]{}3). Optical emission of IGR\,J17098-3628 is 
likely due to X-ray heating of outer regions of the
disk. Scaling the observed $V\sim 20.8$ (see
Fig.\,\ref{image_opt}) to the average absolute magnitude of
LMXBs $M_{\rm V}\simeq1.2$
\citep{vanparadijs81} and correcting for the extinction $A_{\rm V}\sim4.5$
(derived from $N_{\rm H}$) we can estimate the distance to the
source $d\simeq10.5$ kpc.
\section*{Acknowledgments}

This research was supported by the Russian Foundation for Basic
Research (project 05-02-17454), the Presidium of the Russian
Academy of Sciences (the ``Origin and evolution of stars and
galaxies'' program), and the Program of the Russian President
for Support of Leading Scientific Schools (project
NSh-1100.2006.2).

\end{document}